\documentclass[preprint,showpacs,showkeys,preprintnumbers,amsmath,amssym]{revtex4-1}
\usepackage[dvips]{graphicx}
\usepackage{float}
\usepackage{color}
\usepackage{times}
\usepackage{multirow}
\usepackage{hyperref}

\def\KT{Koopmans' theorem}
\def\EA{electron affinity}
\def\UPS{photoelectron spectroscopy}
\def\IPs{ionization potentials}
\def\OMe{methoxy}
\def\PAH{polycyclic aromatic hydrocarbons}
\newcommand{\threelinebrace}{$\left. \begin{array}{c} \\ \\ \\ \end{array} \right\rbrace$}
\newcommand{\twolinebrace}{$\left. \begin{array}{c} \\ \\ \end{array} \right\rbrace$}
\definecolor{pink}{rgb}{1,0,1} 

\begin{document}

\preprint{S.~Shahab et al.}

\title{Theoretical Study of New Acceptor and Donor Molecules based on Polycyclic Aromatic Hydrocarbons}

\author{S.~Shahab Naghavi, Thomas Gruhn, Vajiheh Alijani, Gerhard H. Fecher, Claudia Felser}
\affiliation{Institut f{\"u}r Anorganische Chemie und Analytische Chemie, Johannes Gutenberg - Universtit{\"a}t,  55099 Mainz, Germany}
\email{naghavi@uni-mainz.de}

\author{Katerina Medjanik, Dmytro Kutnyakhov, Sergej A. Nepijko, Gerd Sch{\"o}nhense}
\affiliation{Institut f{\"u}r Physik, Johannes Gutenberg - Universtit{\"a}t, 55099 Mainz, Germany}

\author{Ralph Rieger, Martin Baumgarten, Klaus M{\"u}llen}
\affiliation{Max-Planck-Institut f{\"u}r Polymerforschung, Ackermannweg 10, 55128 Mainz, Germany}

\date{\today}
\begin{abstract}
Functionalized polcyclic aromatic hydrocarbons (PAHs) are an 
interesting class of molecules in which the electronic state 
of the graphene-like hydrocarbon part is tuned by the functional 
group. Searching for new types of donor and acceptor molecules, 
a set of new PAHs has recently been investigated experimentally 
using ultraviolet photoelectron spectroscopy (UPS). In this work, 
the electronic structure of the PAHs is studied numerically with 
the help of B3LYP hybrid density functionals. Using the $\Delta$SCF 
method, electron binding energies have been determined which affirm, 
specify and complement the UPS data. Symmetry properties of molecular 
orbitals are analyzed for a categorization and an estimate of the 
related signal strength. While $\sigma$-like orbitals are difficult 
to detect in UPS spectra of condensed film, calculation provides 
a detailed insight into the hidden parts of the electronic structure 
of donor and acceptor molecules. In addition, a diffuse basis set 
(6-311++G**) was used to calculate \EA ~and LUMO eigenvalues. The 
calculated electron affinity (EA) provides a classification of the 
donor/acceptor properties of the studied molecules. Coronene-hexaone 
shows a high EA, comparable to TCNQ, which is a well-known classical 
acceptor. Calculated HOMO-LUMO gaps using the related eigenvalues 
have a good agreement with the experimental lowest excitation 
energies. TD-DFT also accurately predicts the measured optical gap.
\end{abstract}
\maketitle
\section{Introduction}
\label{Intro}
\begin{figure}[H]
\begin{center}
\includegraphics[width=0.8\linewidth]{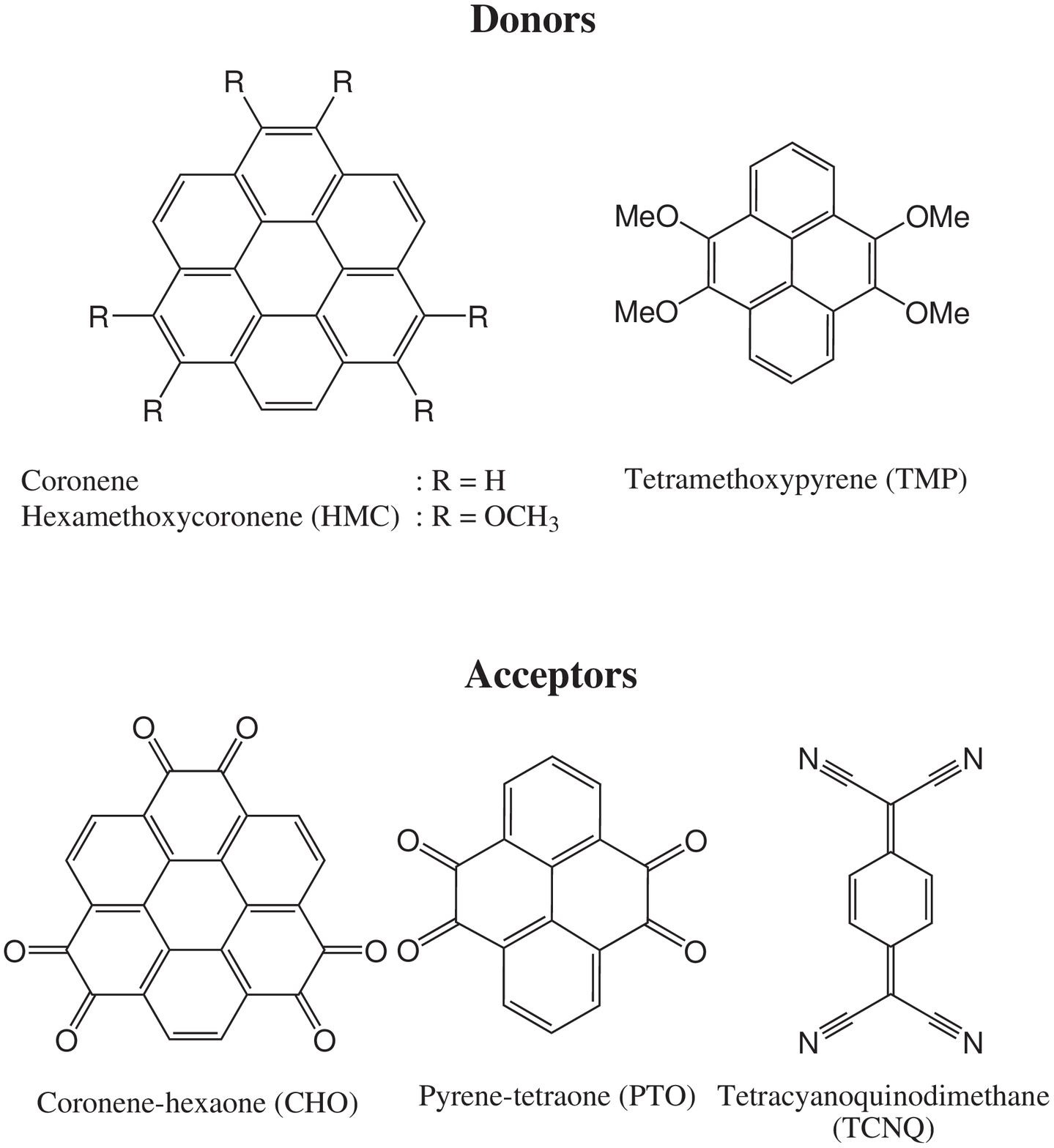}
\caption{Structure of the donor and acceptor molecules studied in this work.}
\label{structure}
\end{center}
\end {figure}

Charge transfer salts have become of particular interest during the 
past two decades, in which both the interest in electronics on the 
molecular scale as well as the methods of chemical synthesis has 
strongly increased. Typically, charge transfer (CT) complexes, 
consisting of $\pi$-electron donors (D) and acceptors (A), show one 
or several properties like metallic conductivity, superconductivity, 
and magnetism that are relevant for nanoelectronic 
applications~\cite{Molecular-Magnet_cht,Organic-superconductors_cht}.
Many of the organic conductors and superconductors are based on the 
bis(ethylenedithio)tetrathiafulvalene (BEDT-TTF) donor and various
acceptors~\cite{Organic-superconductors_cht,Organic-superconductor2_cht}.
Furthermore, poly cyano derivatives have mainly been used as acceptors 
like tetracyanoquinodimethane (TCNQ)~\cite{Organic-conductive_cht}.
Recently, a novel type of donor and acceptor molecules based on 
polycyclic aromatic hydrocarbons (PAHs) has been investigated~\cite{Katherina_cht}.
In this article, a detailed numerical study of the electronic 
properties of new PAHs is presented, which allows to identify 
the experimental results and complements the picture of 
experimentally determined orbital states. \\
\indent Flat aromatic molecules as polycyclic aromatic hydrocarbons 
(PAHs), have attracted attention because of their extended 
$\pi$-electron systems~\cite{Mullen-carrier_cht,Charge-carrier_cht}.
Garito and Heeger concluded that the extension of the $\pi$ system 
would lead to a lowering of intramolecular Coulomb repulsion in the 
anions of the acceptors, resulting in more stable radical anions and 
in highly conducting CT complexes~\cite{pi-extention-better_cht}.
Larger PAH molecules called nanographene have been synthesized in 
the bottom-up approach with desired size, structure, and 
symmetry~\cite{Mullen-graphen-rev_cht}. The electronic structure of 
PAHs can be tuned via the type and number of substitutions in 
periphery. This way the functionalized PAHs may serve as a 
prototype for the creation of a new class of molecules with tailored 
chemical and electronic properties. However, although the electronic 
structure of different donor and acceptor molecules have been widely 
studied, little attention has been paid by theory and experiment to 
illustrate the electronic structure and the effect of functional groups 
on donors and acceptors based on PAHs. \\
\indent In this study we calculate the electronic structure of new acceptors 
and donors based on coronene and pyrene (see Fig.~\ref{structure}).
A hybrid density functional (B3LYP) method is used to calculate the 
ionization potentials (IPs). The results are compared with ultraviolet 
photoelectron spectroscopy (UPS) measurements~\cite{Katherina_cht}.
The comparison allows to assign the measured signals to specific 
molecular orbitals, providing a self consistent validation of the 
experimental results. Furthermore, the calculations provide the IPs 
of orbitals that could not be measured, experimentally.\\
\indent Unfortunately, gas phase measurements were not possible in 
the experiment because of the small amount of material available 
(a few milligrams). Instead, UPS measurements had been performed 
on PAH multilayers on a Au surface. As the molecules are 
preferentially planar to the surface while measurements are 
made in the perpendicular direction $\sigma$-type orbitals 
could not be measured in practice. In the numerical study all 
orbitals can be investigated so that the respective IPs become 
accessible. In general, the IPs of PAHs adsorbed on a metal 
surface are shifted by a constant energy value with respect 
to the gas phase results. The discrepancy stems from the 
photo-hole screening induced by the image charge in the 
metal~\cite{Katherina_cht}. A similar effect occurs in 
multilayer films in which the screening is induced by the 
neighboring molecules. The energy shift between the 
theoretical results that corresponds to isolated molecules 
or the dilute gas phase and the measurements on the condensed 
multilayers can be taken into account by adjusting the HOMO 
levels of both systems. Applying this energy shift to all 
IPs must provide a good correspondence between the experimentally 
measurable orbitals and the theoretical results, which serves 
as a useful consistency check of both measurement and calculation. 
For the given set of molecules the energy shift is very small so 
that the IP results for isolated molecules and those in the 
multilayer can be compared directly.\\
\indent Coronene has a high symmetry (D$_{6h}$), a small band gap 
and the right size for processing techniques. Therefore, 
coronene and its derivatives are promising for applications 
like molecular electronics optoelectronics or 
sensors~\cite{Mullen4_cht,Mullen-AD_cht,Suzuki2006,Casagrande}.
Kato \textit{et al}.~\cite{Coronene-super_cht} also predicted 
superconductivity of coronene in a charge transfer configuration.
In this article, properties of hexamethoxycoronene (HMC) and 
coronene-hexaone (CHO) are studied. The functional groups 
are tailored such that HMC is a donor while CHO is an acceptor 
molecule. Furthermore, calculations of tetramethoxypyrene (TMP) 
and pyrene-tetraone (PTO) have been performed and compared with 
experimental results. These poly aromatic molecules have a 
different symmetry and structure in comparison with  coronene 
based molecules, which makes a comparison of these two families 
instructive. Calculation of tetracyanoquinodimethane (TCNQ), 
a well-known classical acceptor, is used as a benchmark for 
the calculations of the acceptor molecules. All investigated 
molecules are shown in Figure~\ref{structure}. Electronic structures 
of the donor and acceptor molecules are theoretically studied by 
B3LYP/6-31G(d) level of theory and compared with the ultraviolet 
photoemission spectroscopy (UPS) measurements that were carried 
out in parallel~\cite{Katherina_cht}. Electron binding energies 
or ionization potentials of molecules are calculated with the 
delta self consistent field ($\Delta$SCF) method using the B3LYP 
hybrid functional.
\section{Computational details}
\label{Comp}
The calculations were performed using the Gaussian03 
code~\cite{G03} and visualized by VMD~\cite{VMD}.
Calculations were carried out with the B3LYP hybrid 
functional~\cite{B3LYP}. This combination of functionals 
combine the Becke's three parameter exchange potential~\cite{Becke,Becke2}
with the correlation functionals of Lee-Yang-Parr~\cite{LYP,LYP1} 
and Vosko-Wilk-Nusair~\cite{VWN}. B3LYP is one of the 
often employed hybrid functionals~\cite{B3LYP-BEST-for-MOLECULE_cht} 
used in theoretical studies of molecules. \\
\indent The standard 6-31G(d) basis set~\cite{6-31G*} is employed 
for geometry optimizations and energy calculations. This 
level of theory ensures a reliable estimate of structural, 
energetic and spectroscopic parameters for PAHs,  which 
requires an extremely high computational effort for the 
large molecules. \\
\indent Electron binding energies or ionization potentials of 
molecules are calculated by the delta self-consistent 
field ($\Delta$SCF) method applied together with the 
B3LYP hybrid functional. Vertical ionization potentials 
are calculated using $\Delta$SCF method in order to compare 
with UPS data in which vertical ionization potentials
(IP$_{v}$) are measured. In IP$_{v}$ final and  ground 
state have the same geometry. First vertical ionization 
potentials (IP$_{v}$,$_{1}$) are calculated as the 
difference in total energy between a molecule and its 
radical cation at the former geometry. For the higher 
ionization energies (IP$_{v},_{1+n}$), the energy 
difference ($\Delta$ E) between IP$_{v,1}$ and the 
HOMO energy is added to the energy of next higher 
orbitals $\varepsilon_{HOMO-n}$ as a uniform 
shift~\cite{Delta-SCF-DIMER-B3LYP_cht,Delta-SCF_cht,Kohnsham-Orbital2_cht}.
\begin{eqnarray}
{IP_{v,1}}-|\varepsilon_{HOMO}|=\Delta E \\
\label{scf}
|\varepsilon_{HOMO-n}| +\Delta E = IP_{v},_{1+n}
\end{eqnarray}
According to Eq.~\ref{scf} the eigenvalue of the HOMO-n  
orbital should be used in order to calculate the electron 
binding energy (BE) of HOMO-n. A standard view expressed 
in the literature for Kohn-Sham (KS) orbital energies of 
the occupied molecular orbitals lower than HOMO, is that 
they are merely auxiliary quantities.
However, several authors have pointed to the interpretative 
power of KS orbitals in traditional qualitative molecular 
orbital schemes~\cite{KS-1_cht,Kohnsham-Orbital2_cht,
KS-2_cht,Kohnsham-Orbital_cht}, and KS orbitals also show 
a uniform shift with respect to ionization 
energies~\cite{Kohnsham-Orbital2_cht,B3LYP-6-31+g(d)-Ip-Ea-HOMO-LUMO_cht}.
In this sense, B3LYP with its exact Hartree-Fock exchange 
contribution leads to better description of ionization 
energies by reducing the related shift~\cite{Kohnsham-Orbital_cht}. \\
\indent The B3LYP/6-311++G(d,p) level of theory was implemented 
on both the anion and the neutral species for the electron 
affinity calculations. In case of an anionic final state, 
the usage of diffuse function is crucial to ensure the 
accuracy of the calculation. The spin unrestricted B3LYP 
(UB3LYP) is employed in which species are ionized with an
open-shell doublet (one unpaired electron) electron 
configuration. Spin contamination is found in accepted 
limits for the open shell systems, with $<$S$^{2}$$>$
values of about 0.75-0.78 in all cases. At the stationary 
points a vibrational frequency calculation is performed 
for each numerically optimized structure to confirm that 
it is located at the minimum of the potential energy surface. \\
\indent The results obtained with the $\Delta SCF$ method are 
compared with Hartree-Fock calculations based on the 
traditional Koopmans' theorem approach~\cite{Koopman_cht}.
According to \KT, ~the ionization potential (IP) is 
the sign-reversed orbital energy of Hartree-Fock 
eigenvalues: IP $\approx$ -E$_{HOMO}$. Although KT is 
based on some simplification, it shows a reasonable 
agreement with experimental data in several 
cases~\cite{B3LYP-Goodgap_cht}. \KT ~neglects the 
relaxation effect by using the frozen-orbital 
approximation. However, this error is frequently 
compensated by the oppositely directed error due to 
the electron correlation effect, neglected in the 
Hartree-Fock (HF) method. Therefore, the \KT ~is a 
crude but useful and fast approach~\cite{B3LYP-Goodgap_cht}. \\
\indent Moreover, eigenvalues based on B3LYP/6-311++G(d,p) 
are used to estimate HOMO-LUMO gaps and LUMO energies. 
HOMO-LUMO gap based on the energy difference 
between the HOMO and LUMO eigenvalues are compared with
the measured optical gaps.
TD-B3LYP/6-31G(d) is also used to estimate 
the optical band gap (referring to the singlet excited state).
As shown by Zhang \textit{et al} the first excitation 
energy from TD-B3LYP calculations leads to a good 
prediction of the optical gap for most of the DFT 
methods~\cite{LUMO-TDDFT_cht}. Moreover, HOMO-LUMO gaps 
obtained with TD-DFT are not very sensitive to the basis 
set and it is shown that 6-31G(d) can show reasonable 
agreement for optical properties with the experimental 
results for the hydrocarbon molecules~\cite{LUMO-TDDFT_cht,
TDDFT-basis2_cht,TDDFT-basis1_cht}. 
The energy difference 
between the HOMO and LUMO eigenvalues and TD-DFT show the 
similar results and both are in an excellent agreement with 
the experimental results.
\section{Results and Discussion}
The result and discussion section is divided into six 
parts. The first part is about the experimental setup 
in UPS. In the second part, the symmetry of molecules 
and consequences of orbital symmetries on the UPS 
experiment are described. The effect of functional 
groups on coronene and pyrene derivatives is described 
in the third part. In the following two parts, experimental 
and theoretical results of the binding energies for 
coronene and pyrene derivatives are presented. In the sixth 
part, the calculated LUMO and electron affinity (EA) 
values for the studied compounds are discussed.
\subsection{UPS measurements}
Since only small amounts of the new molecules were avaible, 
UPS measurements have not been performed on molecules in the 
gas phase but on UHV-deposited molecules forming some 
multilayers on a Au surface. In the measurement, the work 
function of a clean Au surface increases significantly if the 
acceptor TCNQ or coronene-hexaone is adsorbed~\cite{Katherina_cht}.
It can be understood in terms of a partial charge transfer 
from the metal surface to the first molecular layer thus forming 
the interface dipole with its negative side pointing toward the
organic film. Opposite behavior is observed for the donors HMC, 
TMP and coronene. The decrease of the work function of about 
0.5\,eV indicates the Pauli push-back effect leading to a push-back of
the split-out electrons and thus to a reduction of the surface dipole 
of the metal. In practice, due to charge transfer between molecules and
surface, the polar metal-organic interfaces change the work function 
and thus the binding energies with respect to vacuum level. \\
\indent The binding energies in condensed films of the studied acceptors 
and donors (Fig.~\ref{structure}) are measured by UPS~\cite{Katherina_cht}.
In the UPS spectra the signals derived from the occupied frontier
orbitals of the organic overlayer are referenced to the Fermi 
energy of the Au substrate. By adding the value for the work 
function of the molecular film, the binding energies are 
referenced to the vacuum level. An acceptor molecule in direct 
contact with a metal surface may form a negatively charged species 
as was found for TCNQ on Ni(111)~\cite{Ni-surface_cht}. TCNQ on 
Au, on the other hand, was found to be almost neutral~\cite{Au-surface_cht} 
because its acceptor character does not suffice to form a strong 
charge-transfer complex with Au, which is the  metal with the 
highest Pauling electronegativity. For coronene on a Au surface 
an ionization potential of IP=7.1\,eV is measured (see Table~\ref{coronene})
which is very close to the reported value of 7.29\,eV for coronene 
in the gas phase~\cite{IP-Coronene-gas_cht}. As a result, one has 
that the Au surface shows neutrality with respect to both donor 
(coronene) and acceptor (TCNQ) molecules and experimental and 
theoretical IP values can be compard without considering a large 
energy shift. \\
\indent The following work function values have been measured for films 
with a thickness of several layers deposited on Au: coronene-hexaone 
(5.6\,eV), hexamethoxycoronene (4.8\,eV), coronene (4.8\,eV), 
tetramethoxypyrene (4.9\,eV) and TCNQ (5.9\,eV). \\
\indent Electron binding energies for the molecular orbitals (MOs) can
be extracted from the UPS spectra by assigning peak positions of the
spectra, but one by one assignment for all UPS signals is not possible
since in some cases they result from superposition of several
occupied molecular orbitals with closely spaced binding energies. 
However, in most cases the HOMO signal is well separated from the 
deeper-lying levels and can be evaluated with high precision.
\subsection{Symmetry analysis}
\begin{figure}[H]
\begin{center}
\includegraphics[width=1\linewidth]{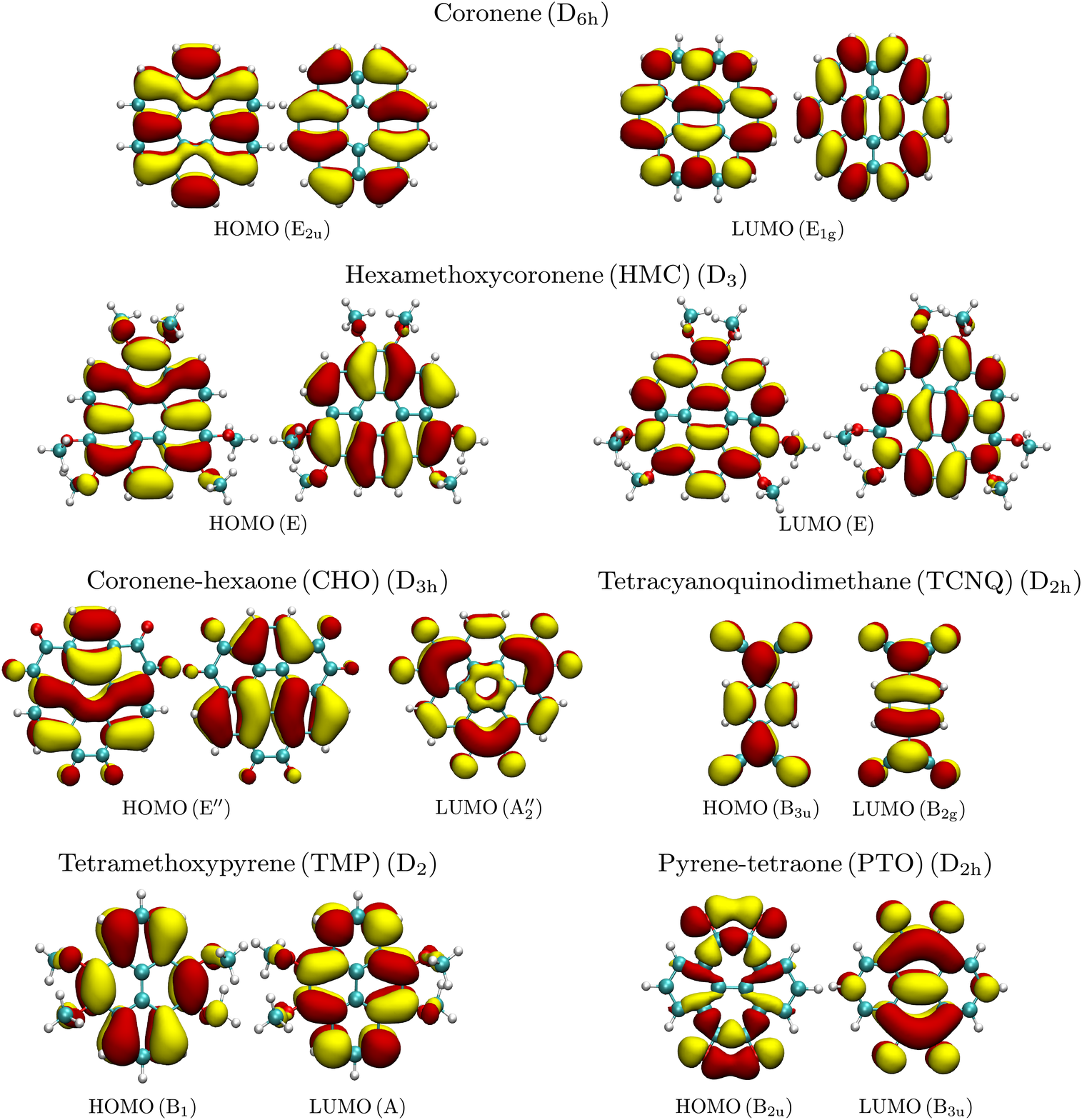}
\caption{HOMO and LUMO orbitals computed for the studied molecules. The symmetry of 
each molecule and the symmetry of the related molecular orbitals are shown in parentheses. Among 
all PAHs donor and acceptor molecules just PTO has a $\sigma$-type HOMO orbital. All 
the coronene derived molecules show a doubly degenerate HOMO.}
\label{surface}
\end{center}
\end {figure}

For the analysis and comparison of UPS data with the calculated 
ionization potentials, the symmetry of the molecular orbitals must 
be extracted. $\sigma$  or $\pi$ symmetry is determined, followed by 
related degeneracy (E, A, ...).This information is shown in 
Figure~\ref{surface} for the frontier orbitals and in Table~\ref{coronene} 
and~\ref{pyrene} for the lower lying orbitals of studied molecules. \\
\indent Having $\pi$ symmetry is essential in order to observe the orbital in 
UPS. The oxygen-derived $\sigma$ states appear only weakly in \UPS ~because 
of the lower cross section of O-2\textit{p} in comparison with carbon 
$\pi$ at the photon energy (21.2\,eV) used in the 
experiment~\cite{Cross-section-O2_cht,Cross-section-O_cht}.
Another reason for the weak $\sigma$-derived signals in the experiment 
might be that the molecules are predominantly oriented parallel to 
the surface and the UPS spectrometer probes the normal emission direction.
Electronic structure calculations allow to determine the electron 
binding energies of the $\sigma$-type orbitals, which are not accessible 
in the experiment. This way a detailed picture of the electronic structures 
of donor and acceptor molecules is available. Orbital symmetries with 
$\sigma$ and $\pi$ character of MOs are listed in Tables~\ref{coronene}, 
\ref{pyrene} and Figure~\ref{surface}. \\
\indent Coronene and its derivatives have a degeneracy of two for the HOMO 
and all orbitals with E symmetry because of more than a 2-fold axis 
(coronene has 6-fold and HMC and CHO have 3-fold axis). CHO and HMC 
have a doubly degenerated HOMO with a $\pi$ characteristic as depicted 
in Figure~\ref{surface}. For HMC and TMP there is no symmetry plane, 
so $\sigma$ and $\pi$ orbitals are not clearly distinguishable for 
the lower lying orbitals. However, they are reported according to 
major contribution of $\sigma$- or $\pi$-type symmetry in the orbital's 
notations in Tables~\ref{coronene} and~\ref{pyrene}. \\
\indent Pyrene derivatives and TCNQ have a non-degenerate orbital because 
of the 2-fold axis. Pyrene-tetraone (PTO) has an exceptional situation 
because HOMO and HOMO-1 have $\sigma$-type symmetry and are in-plane 
as shown in Figure~\ref{structure}. Therefore, they are not traceable 
in UPS because of the low cross section of O-2\textit{p} since these 
MOs are located mainly on the oxygen atoms (see Figure~\ref{surface}).
For PTO, reliable UPS binding energies could not be determined due 
to charging. \\
\indent To sum up, a survey of Tables~\ref{coronene} and~\ref{pyrene} reveals 
that the presence of O atoms and cyanid groups increases the $\sigma$-type 
part of the occupied orbitals of acceptor molecules. Thus, acceptors 
have more  $\sigma$-symmetry orbitals near the HOMO. The experimentally 
inaccessible $\sigma$-type orbitals are provided by the electron structure 
calculations as shown in Tables~\ref{coronene},~\ref{pyrene}.
\subsection{Effect of functional groups}
The electron-withdrawing carbonyl groups of CHO and PTO reduce the 
charge density in the $\pi$ system, while in HMC and TMP, electrons 
are donated to the $\pi$ system by the methoxy (OMe) groups. Therefore, 
the aromatic rings in CHO and PTO are positively charged, while the 
methoxy substituted molecules are negatively charged. Thus, CHO is 
an acceptor in comparison with coronene and HMC, and PTO is an
acceptor compared to TMP. \\
\indent The bond lengths of CHO  along the symmetry plane are distinctly 
larger than those of coronene. The first C--C bond length at the
edge of the double bonded oxygen atoms increases from 1.37\,{\AA} 
(coronene) to 1.54\,{\AA} for CHO. The other bond lengths increase 
by about 0.02\,{\AA} with respect to coronene. For CHO, all the bond 
lengths inside the aromatic ring are larger than those in HMC and 
coronene. In HMC, the C--C bond lengths are of the same size as 
those of coronene, but Mulliken population shows that the C atoms 
of the aromatic rings are slightly more charged compared to coronene. 
In principle, the similar behavior is found for TMP and PTO. \\
\indent Donor and acceptor molecules are best investigated by analyzing 
the molecular orbitals. Any change in the functional groups of 
the molecules will be reflected in the binding energies and the 
molecular orbital energies. Therefore, evaluation of molecular 
orbitals shows how the functional groups tune the electronic 
structure. Also, with the help of ab-initio calculations one 
can analyze details of the electronic structure and in particular 
$\sigma$ orbitals which are not accessible by UPS measurement.
Also underlying orbital structure of overlapping UPS signals 
can be analyzed by the calculation.
\subsection{Electron binding energies of coronene derivatives}
Theoretical and experimental binding energies of coronene and its 
derivatives coronene-hexaone (CHO) and hexamethoxycoronene (HMC) 
are listed in Table~\ref{coronene}. A survey of Table~\ref{coronene} 
reveals that vertical electron binding energies (IPs) calculated 
in $\Delta$SCF-B3LYP approximation are, on the whole, in good 
agreement with the corresponding experimental IPs. Calculations 
based on the $\Delta$SCF method correspond to the vertical IP of 
the molecules in the gas phase. In general, they may differ from 
measurements of condensed layers on a metal surface by a constant 
energy value. As shown in Table~\ref{coronene} the energy shift 
between the multilayer measurements and the calculations corresponding 
to the gas phase are very small, which corresponds to the fact that 
the IP=7.1\,eV of coronene on Au is close to the reported gas phase 
value IP=7.29\,eV~\cite{IP-Coronene-gas_cht}.  \\
\indent For coronene and the derivatives the Koopmans' energies for the HOMO 
differ by less than 0.5\,eV from the $\Delta$SCF binding energies. For 
the deeper-lying orbitals, the Koopmans' energies are too large in 
comparison with $\Delta$SCF and experimental results. This deviation is 
caused by correlation and relaxation effects which are neglected in
Koopmans' theorem. For comparison, results from Koopmans' theorem are 
presented for HMC in Table~\ref{coronene}. For coronene and CHO deviations 
of the Koopmans' energies are of the same order of magnitude. Note that 
the $\sigma$-like orbitals of CHO appear only very weakly in UPS spectra. 
$\Delta$SCF-B3LYP results can cover a wide range of MOs and show good 
agreement with the UPS data. \\
\begin{table*}[ht!]
\caption{Electron binding energies (eV) for coronene and its derivatives with the symmetry
labels calculated by B3LYP/6-31G* and compared to vertical excitation energies from ultra violet \UPS ~(UPS) and \KT ~(KT).}
\begin{ruledtabular}
\begin{tabular}{l c c c c c c c r c c c r}
& \multicolumn{3}{c}{CHO(D$_{3h}$)}  & \multicolumn{4}{c}{HMC(D$_{3}$)} &\multicolumn{3}{c}{Coronene(D$_{6h}$)}\\
\cline{2-4}
\cline{5-8}
\cline{9-11}
MOs & Symm. & $\Delta$SCF & UPS & Symm. & KT & $\Delta$SCF & UPS & Symm. & $\Delta$SCF & UPS  \\
\hline
HOMO    & E$^{\prime\prime}$($\pi$)     & 8.78  & 8.8   & E($\pi$) & 6.86  & 6.41  & 6.5   & E$_{2u}$($\pi$)    & 6.84  & 7.1 \\
HOMO-1  & E$^{\prime}$($\sigma$)        & 8.80  & ..... & E($\pi$) & 8.74  & 7.66  & \multirow{3}{*}{\threelinebrace 7.9\quad\;} & E$_{1g}$($\pi$)    & 8.11  & \multirow{3}{*}{\threelinebrace 8.8\quad\;} \\
HOMO-2  & A$_{1}^{\prime}$($\sigma$)    & 8.82  & ..... & A$_{2}$($\pi$)  & 9.25  & 7.80  &       & B$_{2g}$($\pi$)    & 8.56  \\
HOMO-3  & A$_{1}^{\prime\prime}$($\pi$) & 9.23  & ..... & A$_{1}$($\pi$)  & 9.44  & 8.15  &       & B$_{1g}$($\pi$)    & 8.62  \\
HOMO-4  & E$^{\prime}$($\sigma$)        & 10.05 & ..... & E($\sigma$)        & 11.03 & 8.38  & ..... & A$_{2u}$($\pi$)    & 9.84  & ..... \\
HOMO-5  & A$_{2}^{\prime}$($\sigma$)    & 10.11 & ..... & A$_{1}$($\sigma$)  & 11.59 & 8.51  & ..... & E$_{2g}$($\sigma$) & 9.94  & ..... \\
HOMO-6  & E$^{\prime\prime}$($\pi$)     & 10.33 & \multirow{2}{*}{\twolinebrace 10.3\quad\;\;} & E($\pi$) & 11.65 & 9.11  & \multirow{3}{*}{\threelinebrace 9.3\quad\;} & E$_{2u}$($\pi$)    & 10.25 & ..... \\
HOMO-7  & A$_{2}^{\prime\prime}$($\pi$) & 10.70 &       & A$_{2}$($\pi$)  & 11.65 & 9.17  &       & B$_{2u}$($\sigma$) & 11.09 & ..... \\
HOMO-8  & E$^{\prime\prime}$($\pi$)     & 11.80 & ..... & A$_{2}$($\pi$)  & 12.47 & 9.52  &       & E$_{1u}$($\sigma$) & 11.13 & ..... \\
HOMO-9  & E$^{\prime}$($\sigma$)        & 11.84 & ..... & E($\pi$)        & 12.64 & 9.71  & ..... & A$_{2g}$($\sigma$) & 11.54 & ..... \\
HOMO-10 & A$_{1}^{\prime\prime}$($\pi$) & 12.37 & ..... & E($\pi$)        & 12.94 & 9.95  & ..... & B$_{1u}$($\sigma$) & 11.55 & ..... \\
\end{tabular}
\end{ruledtabular}
\label{coronene}
\end{table*}
\indent The first ionization potential (binding energy of HOMO) of HMC is 
shifted up by 0.5\,eV in comparison to coronene. As displayed in
Table~\ref{coronene}, hexamethoxycoronene has three main signals 
in the experimental spectrum below 10\,eV~\cite{Katherina_cht}. The 
first of them at 6.5\,eV is the HOMO with $\pi$-like symmetry with 
the double degenerate representation E (Figure~\ref{surface}). 
The calculation yields a binding energy ($\Delta$-SCF) of 6.41\,eV 
in excellent agreement with the photoelectron spectra. The calculated 
binding energy of electrons in the HOMO of HMC is about 0.4\,eV lower 
than that of coronene due to the six methoxy groups on the periphery 
which increase the charge density in the $\pi$ system. The effect of 
the methoxy groups is weaker than that of electron-withdrawing keto 
groups of CHO since the methoxy groups contain the electron-donor 
methyl and the acceptor oxygen which partially compensate each other.
There are three pairs of \OMe ~groups in the ring system and each 
pair is in trans configuration (see Figure~\ref{structure} and~\ref{surface}).
The \OMe~groups can easily rotate in a solvent but this rotation 
will be blocked on the surface. The orientation of the methoxy group 
configurations has a strong influence also on the interaction between 
metal surface and molecule. The trans configuration of \OMe~groups may 
reduce the contact of the aromatic ring system with the surface so 
that it can effectively suppress charge transfer between molecule and 
metal surface. The second signal in the UPS spectra appears at 7.9\,eV 
and is assigned to the group  HOMO-1, HOMO-2 and HOMO-3 all having 
$\pi$-like symmetry. Note that there is no strict $\pi$- and $\sigma$-type 
symmetry in HMC, but we can identify '$\pi$-like' and '$\sigma$-like'
symmetry by comparison with the parent molecule coronene. HOMO-4 and 
HOMO-5 are very weak in the experimental spectra because they have 
$\sigma$-like symmetry, and contain a high contribution from oxygen 
2$p$ which has a low cross section at the photon energy (21.2\,eV) used 
in the experiment~\cite{Cross-section-O2_cht,Cross-section-O_cht}. Another 
reason for the weak $\sigma$-derived signals in the experiment might be 
that the molecules are predominantly oriented parallel to the surface and 
the UPS spectrometer probes the normal emission direction. For perfectly 
parallel orientation, $\sigma$ emission is forbidden by symmetry selection 
rules. The UPS signal at 9.3\,eV corresponds to the group of the HOMO-6 to 
HOMO-8 states with $\pi$-like symmetry.  \\
\indent \textbf{Coronene-hexaone} (CHO) constitutes a more complicated adsorption 
behavior than HMC. CHO is a planar molecule with high electron affinity. 
The first layer being in direct contact with surface experiences a substantial 
charge transfer from Au showing up in terms of a new interface state~\cite{Katherina_cht}.
For comparison with theory we refer the UPS data for higher CHO coverage, 
where the interface state has disappeared in the spectra. In the multilayer 
coverage two signals rise in the region below 11\,eV. As shown in 
Table~\ref{coronene}, the first signal corresponds to the HOMO and appears 
at 8.8\,eV in very good agreement with the $\Delta$SCF calculation (8.78\,eV). 
This value is similar to TCNQ which is a strong classical acceptor(Figure~\ref{pyrene}). 
The binding energy of electrons in the HOMO of CHO has increased by more 
than 2\,eV with respect to coronene because of the electronegative keto
groups.
HOMO-1 and HOMO-2 as well as HOMO-4 and HOMO-5 are weak signals in the 
experimental spectra since they are derived from the oxygen 2p$_{x,y}$ 
orbitals that are aligned in the molecular plane. HOMO-6 and HOMO-7 
($\pi$ symmetry) constitute the second UPS signal at 10.3\,eV in the
measured spectra. The $\Delta$SCF calculation again shows good agreement.
Obviously the six carbonyl groups strongly alter the electronic structure 
of CHO compared to unfunctionalized coronene, and make CHO a strong acceptor. 
\subsection{Binding energies of pyrene derivatives}
For \textbf{tetramethoxypyrene} (TMP) and \textbf{tetracyanoquinodimethane} 
(TCNQ), the theoretical binding energies are smaller than the experimental 
values. For the HOMO of TMP the difference is 0.42\,eV, similar as for the 
overlapping HOMO-1 and HOMO-2 signals. HOMO-3 to HOMO-4 and HOMO-9 to HOMO-10 
have $\sigma$-like symmetry and appears weakly in the experimental spectra 
due to a large oxygen 2p content. The $\pi$-like  group HOMO-6 to HOMO-8 
can not be resolved experimentally and shows up as one signal centered at 9.2\,eV. \\
\begin{table*}[ht!]
\caption{Same as table~\ref{coronene}, but for pyrene derivatives and TCNQ.}
\begin{ruledtabular}
\begin{tabular}{l c c c c c c c c c}
& \multicolumn{3}{c}{PTO(D$_{2h}$)}  & \multicolumn{3}{c}{TMP(D$_{2}$)} &\multicolumn{3}{c}{TCNQ(D$_{2h}$)}\\
\cline{2-4}
\cline{5-7}
\cline{8-10}
MOs & Symm. & $\Delta$SCF & UPS & Symm. & $\Delta$SCF & UPS & Symm. & $\Delta$SCF & UPS  \\
\hline
HOMO    & B$_{2u}$($\sigma$)  & 8.70  & ..... & B$_{1}$($\pi$) & 6.18  & 6.7   & B$_{3u}$($\pi$) &  8.87 & 9.3 \\
HOMO-1  & A$_{g}$($\sigma$)   & 8.90  & ..... & B$_{3}$($\pi$) & 7.09  & \multirow{2}{*}{\twolinebrace 7.7\quad\;} & B$_{1g}$($\pi$) & 10.49 &    \multirow{2}{*}{\twolinebrace 10.8\quad\;\;} \\
HOMO-2  & B$_{2g}$($\pi$)     & 8.92  & ..... & B$_{2}$($\pi$) & 7.30  &       & B$_{2g}$($\pi$) & 10.91  \\
HOMO-3  & A$_{u}$($\pi$)      & 9.50  & ..... & A ($\pi$)      & 8.03  & ..... & B$_{3g}$($\sigma$) & 11.56 & ..... \\
HOMO-4  & B$_{1g}$($\pi$)     & 9.67  & ..... & B$_{3}$($\sigma$) & 8.06  & ..... & B$_{2u}$($\sigma$) & 11.58 & ..... \\
HOMO-5  & B$_{3g}$($\sigma$)  & 9.96  & ..... & A($\pi$)      & 8.28  & \multirow{5}{*}{\threelinebrace 9.3\quad\;}      & A$_{u}$($\sigma$)  & 11.87 & ..... \\
HOMO-6  & B$_{1u}$($\sigma$)  & 10.26 & ..... & B$_{1}$($\pi$) & 8.66  &       & B$_{1g}$($\pi$) & 11.88 & ..... \\
HOMO-7  & B$_{3u}$($\pi$)     & 10.60 & ..... & B$_{2}$($\pi$) & 8.72  &       & B$_{1u}$($\pi$) & 11.89 & ..... \\
HOMO-8  & B$_{2g}$($\pi$)     & 11.66 & ..... & B$_{2}$($\pi$) & 8.96  &       & A$_{g}$($\sigma$)  & 12.01 & ..... \\
HOMO-9  & B$_{3g}$($\sigma$)  & 12.08 & ..... & B$_{1}$($\pi$) & 9.42  & ..... & B$_{3g}$($\sigma$) & 12.34 & ..... \\
HOMO-10 & A$_{g}$($\sigma$)   & 12.10 & ..... & B$_{2}$($\sigma$) & 9.71  & ..... & B$_{2u}$($\sigma$) & 12.38 & ..... \\
\end{tabular}
\end{ruledtabular}
\label{pyrene}
\end{table*}

\indent For \textbf{pyrene-tetraone} (PTO) reliable UPS binding energies could 
not be determined due to charging. Moreover, the HOMO has $\sigma$ 
symmetry and is thus not observable in the UPS experiment. Charging was 
also observed for thick CHO films (above 10 nm thickness). For TCNQ, 
HOMO-1 and HOMO-2 agree fairly well with the experimental signal 
centered at 10.6\,eV. HOMO-3 to HOMO-5 have $\sigma$ symmetry and appear 
only weakly in UPS spectra.
\subsection{LUMO and electron affinity}
The strength of an acceptor molecule is measured by its electron affinity 
(EA) which is the energy released when adding one electron to the lowest 
unoccupied molecular orbital. An acceptor must have a high electron affinity, 
while low electron affinity is required for effective donor molecules.
Since EA depends on the lowest unoccupied molecular orbital (LUMO), it
can be calculated more accurately by considering the final state (anion). 
In order to describe the anion with an extra electron sufficiently, a diffuse 
basis set is required. We use the B3LYP/6-311++G(d,p) basis to study donor 
and acceptor molecules. \\
\indent The calculated EA of coronene is 0.48\,eV  which is in good agreement with 
the reported experimental value of 0.470$\pm$0.009\,eV~\cite{EA-Coronene_cht}.
A high electron affinity is the major characteristic of strong acceptors 
like TCNQ. Reported values for the EA of TCNQ vary with the applied measurement 
technique~\cite{Ea-TCNQ-Exp_cht,Ea-Gap-TCNQ-good_cht,Ea-TCNQ-Exp-2_cht,Ea-TCNQ-Science_cht}.
Calculations values of EA are in the range of 3.4$\pm$4\,eV~\cite{Ea-TCNQ-Theo_cht}.
Using B3LYP/6-311++G**, we obtained an EA of 3.62\,eV, while B. Mili\'an et al. 
report a value of 3.42\,eV calculated with B3LYP/cc-pVDZ  and 3.28\,eV calculated 
with CCSD/aug-cc-pVDZ. All these values are close to the reported experimental 
value of EA=3.3$\pm$3~\cite{Ea-TCNQ-Science_cht,Ea-Gap-TCNQ-good_cht}. \\
\indent The electron affinity of CHO is 3.5\,eV (see first column of Table~\ref{EA}) 
which is close to the electron affinity of TCNQ (3.62\,eV). Therefore CHO 
is expected to be a strong electron acceptor. CHO has a large electron 
affinity and an extended $\pi$ system which makes it an interesting 
alternative to common acceptors. \\
\begin{table}[ht!]
\scriptsize
    \caption{$\Delta$SCF electron affinity (EA) using 6-311++G**
 basis set. Moreover the HOMO-LUMO gaps are calculated
using TD-DFT/6-31G* (singlet excitation), second column and B3LYP/6-311++G** (LUMO-HOMO eigenvalue differences) third 
column. All values are given in eV.}
    \centering
        \begin{ruledtabular}
        \begin{tabular}{l c c c c c}
 Molecule & EA & EA & Gap & Gap & Gap\\
          & (B3LYP) & (Exp)  & (TD-DFT) & ($\varepsilon_{LUMO}-\varepsilon_{HOMO}$) & (Exp) \\
        \hline
Coronene              & 0.48 & \textbf{0.47}\footnotemark[1]& 3.99 & 3.23 & 3.54\footnotemark[3]   \\
Coronene-Hexaone      & 3.50 & \ldots & 2.98 & 2.57 & 3.20\footnotemark[3]   \\
Hexamethoxycoronene   & 0.51 & \ldots & 3.81 & 3.09 & 3.30\footnotemark[3]   \\
Pyrene-Tetraone       & 2.48 & \ldots & 3.48 & 2.22 &              \\
Tetramethoxypyrene    & 0.41 & \ldots & 3.67 & 3.58 &              \\
TCNQ                  & 3.62 & \textbf{3.3}\footnotemark[2]& 3.04 & 2.54 & 3.10\footnotemark[4]\\
        \end{tabular}
        \end{ruledtabular}
\tiny
\footnotetext[1]{Ref~\textbf{\cite{EA-Coronene_cht}}}  
\footnotetext[2]{Ref~\textbf{\cite{Ea-TCNQ-Science_cht,Ea-Gap-TCNQ-good_cht}}}  
\footnotetext[3]{Ref~\cite{Mullen-AD_cht}}  
\footnotetext[4]{Ref~\cite{TCNQ-Gap_cht}} 
    \label{EA}
\end{table}
\indent In Table~\ref{EA} the experimental gaps are compared with the theoretical 
gaps. To estimate the optical gap, both TD-DFT and the difference of B3LYP 
HOMO-LUMO eigenvalues are used and the results are listed in the second and 
third column of Table~\ref{EA} respectively. 
In the case of large molecules there is a good agreement 
between the eigenvalue differences and the measured optical gap.
In the earlier study, it is shown that there is a large deviation 
between the calculated gap using B3LYP and experimental 
results~\cite{B3LYP-Goodgap_cht}. In the same paper, 
discrepancy depends to the size of molecule and it 
decreases by increasing the size of hydrocarbons. 
Therefore, one can concludes the good agreement in our calculations 
should be attributed to the large size of the hydrocarbons. 
For all compounds, the calculated HOMO-LUMO eigenvalue differences are 
smaller than the measured gaps. The gaps obtained from TD-DFT are spread 
symmetrically around the experimental values. In general, the calculated gaps 
from TD-DFT and the HOMO-LUMO eigenvalues deviate less than $20\%$ from 
the experimental gaps, which is a reasonable result for this quantity. 
The fair agreement between the experimental values and the results calculated 
with B3LYP/6-31G* indicates that this calculation method may be used also 
for the larger molecules of this family called nanographene for which 
more correlated methods like post Hartree-Fock or even DFT with the 
diffuse basis sets are difficult or not applicable.
\section{Summary}
In this paper, we have reported theoretical and experimental electron 
binding energies of different MOs for coronene, pyrene and their methoxy 
and keton-derivatives as \PAH~ and compared them with TCNQ as a well-known
classical acceptor. The studied compounds demonstrate how the functional 
groups can tune the molecules' electronic state, providing moderate donor 
or strong acceptor properties. \\
\indent Comparison of electron structure calculations and \UPS ~data provides new 
insights into the electronic structure of the new compounds. The calculated 
molecular orbitals have been used to identify and complement the experimental 
results. UPS measurements of the compounds on a Au surface provide IPs in 
an agreement with experimental and calculated values for the gas phase. 
The absence of an energy shift, induced by the metal substrate or the 
neighboring molecules, is so far not understood, but the good agreement 
between the various UPS signals and the calculated $\pi$-type orbitals 
demonstrates the direct comparability of the molecule's IPs in the 
condensed and the gas state. In fact, $\Delta$SCF-B3LYP reproduces the 
whole range of IPs in the measured UPS spectra, while \KT ~shows a good 
agreement just for the first IP. The calculated electron affinities allow 
to rate the donor or acceptor character of the molecules. They confirm 
that CHO is a strong acceptor. Its high \EA ~(3.5\,eV), which characterizes 
a good acceptor, is almost as large as the affinity of TCNQ (3.62\,eV).
HMC, coronene and TMP with their low \IPs~and rich aromatic systems are 
suitable candidates for donors. HMC is an appropriate choice for surface 
applications due to its trans isomerism. 
Especially interesting for the formation of charge transfer complexes are 
the molecules CHO and HMC. Our results show that they are good acceptor 
and donor molecules, respectively, and with their coronene centers they 
have a very similar molecular structure. This is analogous to the TTF-TCNQ 
pair with the benzene core which is studied since a long time. The proposed 
coronene-based donor-acceptor molecules could be the starting point of a 
new generation of charge transfer salts based on molecular nanographenes.
\section{Acknowledgments}
Financial support by the Deutsche Forschungs Gemeinschaft DFG through
the Collaborative Research Center {\it "Condensed Matter Systems with
Variable Many-Body Interactions"}(Transregio SFB/TRR 49), the graduate school of excellence MAINZ as well as the center COMATT 
are gratefully acknowledged.

\bibliography{database_in_org}

\end{document}